\documentclass[12pt]{article}

\usepackage[letterpaper,top=2cm,bottom=2cm,left=2cm,right=2cm,marginparwidth=1.75cm]{geometry}
\usepackage{setspace}
\usepackage{authblk} 

\usepackage{graphicx}
\usepackage{pdflscape}
\usepackage{subcaption}
\usepackage{booktabs}
\usepackage{amsmath}
\usepackage{bm}
\usepackage{tocloft}
\usepackage{blindtext}
\usepackage[numbers]{natbib}
\usepackage{xcolor}
\usepackage[colorlinks=true, allcolors=blue]{hyperref}
\providecommand{\keywords}[1]{\textbf{\textit{Keywords }} #1}

\title{Dynamic Prediction of High-density Generalized Functional Data with Fast Generalized
Functional Principal Component Analysis}
\author[1,*]{Ying Jin}
\author[1]{Andrew Leroux}
\affil[1]{Department of Biostatistics and Informatics, University of Colorado Anschutz Medical Campus, Aurora, CO, USA}
\affil[*]{Corresponding author: Ying Jin, ying.jin@cuanschutz.edu}

\date{}

\begin{document}
\maketitle

\begin{abstract}
Dynamic prediction, which typically refers to the prediction of future outcomes using historical records, is often of interest in biomedical research. For datasets with large sample sizes, high measurement density, and complex correlation structures, traditional methods based on Generalized Linear Mixed Models are often infeasible because of the computational burden associated with \textit{both} data size and model complexity. Moreover, many implementations of mixed-effects models do not directly facilitate out-of-sample predictions for generalized outcomes. To address these issues, we develop a novel approach for dynamic predictions based on a recently developed method estimating complex patterns of variation for exponential family data: fast Generalized Functional Principal Components Analysis (fGFPCA). Our method is able to handle large-scale, high-density repeated measures much more efficiently than existing methods, with its implementation feasible even on personal computational resources (e.g., a standard desktop or laptop computer). The proposed method makes highly flexible and accurate predictions of future trajectories for data that exhibit high degrees of nonlinearity, and allows for out-of-sample predictions to be obtained without reestimating any parameters. A simulation study is designed and implemented to illustrate the advantages of this method. To demonstrate its practical utility, we also conducted a case study to predict diurnal active/inactive patterns using accelerometry data from the National Health and Nutrition Examination Survey (NHANES) 2011-2014. Both the simulation study and the data application demonstrate the better predictive performance and high computational efficiency of the proposed method compared to existing methods. The proposed method also obtains more personalized prediction that improves as more information becomes available, which is an essential goal of dynamic prediction that other methods fail to achieve. 
\end{abstract}


\section{Introduction}
\label{sec:introduction}


Biomedical investigators are often interested in predicting future
observations of subjects based on their historical data, where we hope to make predictions that are both 1) individualized; and 2) improves as new information is collected. Such problems are typically termed dynamic prediction. The two desirable properties of these methods are important to support timely clinical decision-making that is customized to each unique individual. Examples of such problems include predicting the change in certain biomarkers over time \citep{leroux2016}, predicting the future status of physical activity using historical data collected by wearable devices, predicting the development of lesion based on imaging data, et cetera. 

Traditionally, this type of data has been referred to as ``repeated measures" and modeled either marginally, such as generalized estimating
equations, or conditionally on specific subjects, such as mixed-effect
models \citep{davidian2003, Laird1982, liang1986, lindstrom1990}. Subject-specific predictions are often made based on the correlation between repeated measures from the same subject. Examples include \citet{GLMMadaptive}, which makes predictions by fitting a generalized linear mixed model (GLMM) and approximates the posterior distribution of parameters of interest with a multivariate normal distribution and their posterior mean. It adopts the Monte Carlo scheme to efficiently approximate the posterior likelihood. Moreover, such repeated measures are often modeled together with a time-to-event outcome to predict the risk of an event, which is updated with the value of repeated measures \citep{Rizopoulos2017, suresh2017}. Popular model frameworks include joint models and landmarking. The former specifies a joint distribution of the repeated measure and the time-to-event outcome. Random effects are typically introduced to model repeated measures and/or introduce correlation between multiple series of repeated measures. The latter estimates models and makes predictions conditioning on a series of ``landmark" times, although these models typically use time-dependent fixed effects alone for prediction and do not include random effects. \citep{vanHouwelingen2007, Rizopoulos2011}

However, with
the notable exception of \citet{leroux2016}, existing methods are either 
limited in their flexibility (both fixed and random effect structures,
e.g.~random slope and intercept), computationally unfeasible for large
datasets, or both. Moreover, when the repeated measure is non-Gaussian, out-of-sample prediction often requires reestimating part or all of the model parameters, which adds to the computational burden. These issues are particularly acute for densely measured data, as the computational burden increases substantially with increased data size and measurement density, motivating investigators to look for more efficient methods. A statistical framework well-suited for high-density data with complex association structures is known as Functional Data Analysis (FDA) \citep{fdaR2024, fda2005}, which conceptualizes high-density repeated measures as the discrete realization of a continuous latent process or a continuous function measured with error. To better explain and illustrate the methodological framework of the FDA, we use data from the National Health and Nutrition Examination Survey (NHANES) as a motivational example throughout this manuscript. The NHANES data from 2011-2014 includes minute-level physical activities collected with wearable activity monitors, which were then dichotomized into binary activity indicators. Figure \ref{fig:nhanes_data} shows the data of six participants. The black dots indicate the activity status of each minute, with 1 indicating active and 0 inactive. The measurement grid is very dense with 1440 repeated measures over one day. The red lines are the estimated probability curves derived from the observed outcomes, which are interpreted as the \textit{latent} process that specifies the distribution of outcomes at each minute. As the figure reveals, these diurnal trends (and hence, latent functions) are highly nonlinear. Moreover, there is substantial heterogeneity between individuals, which cannot be fully explained using participant characteristics modeled as fixed effects, with unexplained variation continuing to exhibit a high degree of nonlinearity.

Methodology for estimating functional regression models with subject-specific random effects has focused primarily on continuous/Gaussian outcomes, often modeling
subject-specific random effects with Functional Principal Component Analysis (FPCA) \citep{chiou2012, goldberg2014, shang2017, kraus2015}. Existing methods often use partial observations for prediction with an intercept-only model.
\citet{leroux2016} proposed
the Functional Concurrent Regression (FCR) framework, which can incorporate
the effect of subject-specific predictors, accompanied by an R package \emph{fcr} \citep{fcrpkg}. \citet{delaigo2016} achieved
similar goals using Markov Chains to model the transition between discrete realizations of a continuous function, though the model it proposed did not include random effects.

Software and methods for non-Gaussian functional data are currently even more limited. Software is largely limited to the \emph{refund} \citep{refundpkg} and \emph{registr} \cite{Wrobel2018}
packages in the statistical software R, both developed based on the Generalized Functional Principal Component Analysis (GFPCA) framework. Functional
regression models of non-Gaussian outcomes have also been proposed. \citet{chen2013} proposed approaches to fit
marginal functional models that are compatible with multilevel generalized outcomes, although marginal approaches are not suitable for making subject-specific dynamic predictions. \citet{gertheiss2017} identified the distinction between marginal and conditional estimates and pointed out that directly applying a marginal method (e.g. single-level FPCA) to conditional estimation could cause bias. It also proposed to address this problem using a two-stage joint estimation strategy. \citet{goldsmith2015} proposed estimating parameters with Bayesian method in \emph{Stan}, but the model
framework only takes into account the fixed effect of time-invariant
covariates. \citet{linde2019} used an adapted Bayesian
variational algorithm for FPCA of binary and count data. However, most of these methods are not adaptable to dynamic prediction problems, and others can be time consuming in parameter estimation, making it unfeasible for large-scale, high-density datasets, such as the following motivation example of NHANES accelerometer data in Section \ref{sec:data}.

To address the challenges mentioned above, in this paper, we propose a novel method based on the principles of GLMM and GFPCA, which is fast and scalable for large datasets with high-density repeated measures and readily allows for dynamic predictions of new subjects without the need for parameter reestimation. Section \ref{sec:data} introduces the NHANES activity indicator as a motivation example. Section \ref{sec:method} explains the model framework of the proposed method, laying out the algorithm for implementation. In Section \ref{sec:simulation}, we illustrate the performance and efficiency of the proposed method in a simulation study, comparing our method to classic methods used for dynamic prediction of non-Gaussian repeated measures. In Section \ref{sec:case_study},
we demonstrate the practical use of this method in our motivation dataset: the minute-level active/inactive diurnal patterns from the 2011-2014 National Health and Nutrition Examination Survey (NHANES). We conclude with a discussion in Section \ref{sec:discussion}. 

\section{NHANES minute-level activity indicator}
\label{sec:data}

The National Health and Nutrition Examination Survey (NHANES) is a large study conducted by United States Centers for Disease Control and Prevention (CDC). The study focuses on non-institutionalized US population and covers many aspects of public health, including demographic,
socioeconomic, nutrition and
dietary intake, as well as physical fitness and activity. In this paper, we will use the physical activity monitor (PAM) data from 2011-2014 as a motivation example of dynamically predicting generalized functional data. 

The analysis sample includes 8763 participants aged 20 to 80 years old, with an average age of 49.53. 52.05\% (4561)  of the sample are female. The range of BMI is 13.6-82.90 with an average of 29.08. The raw sub-second accelerometry data over a week was summarized at the minute level using the Monitor Independent Movement Summary (MIMS) unit. As a result, a single subject will have 1440 observations over each day. The minute-level MIMS was then dichotomized into minute-level binary activity indicators, where 1 indicates active status and 0 sedentary. The cutoff value of dichotomization was determined to be 10.558, following the suggestions in \citet{preprocess} and \citet{Koster2016}. The threshold value was developed based on its association with other activity measures. Finally, active/inactive status at each minute is summarized across days using the median value, rounded up in the event of a non-integer median. Ultimately, the preprocessing results in 1440 binary observations per participant.  \citet{leroux2022} provides additional details on data preprocessing steps. 

\begin{figure}
\centering

\begin{subfigure}{0.9\textwidth}
\includegraphics[width=\textwidth]{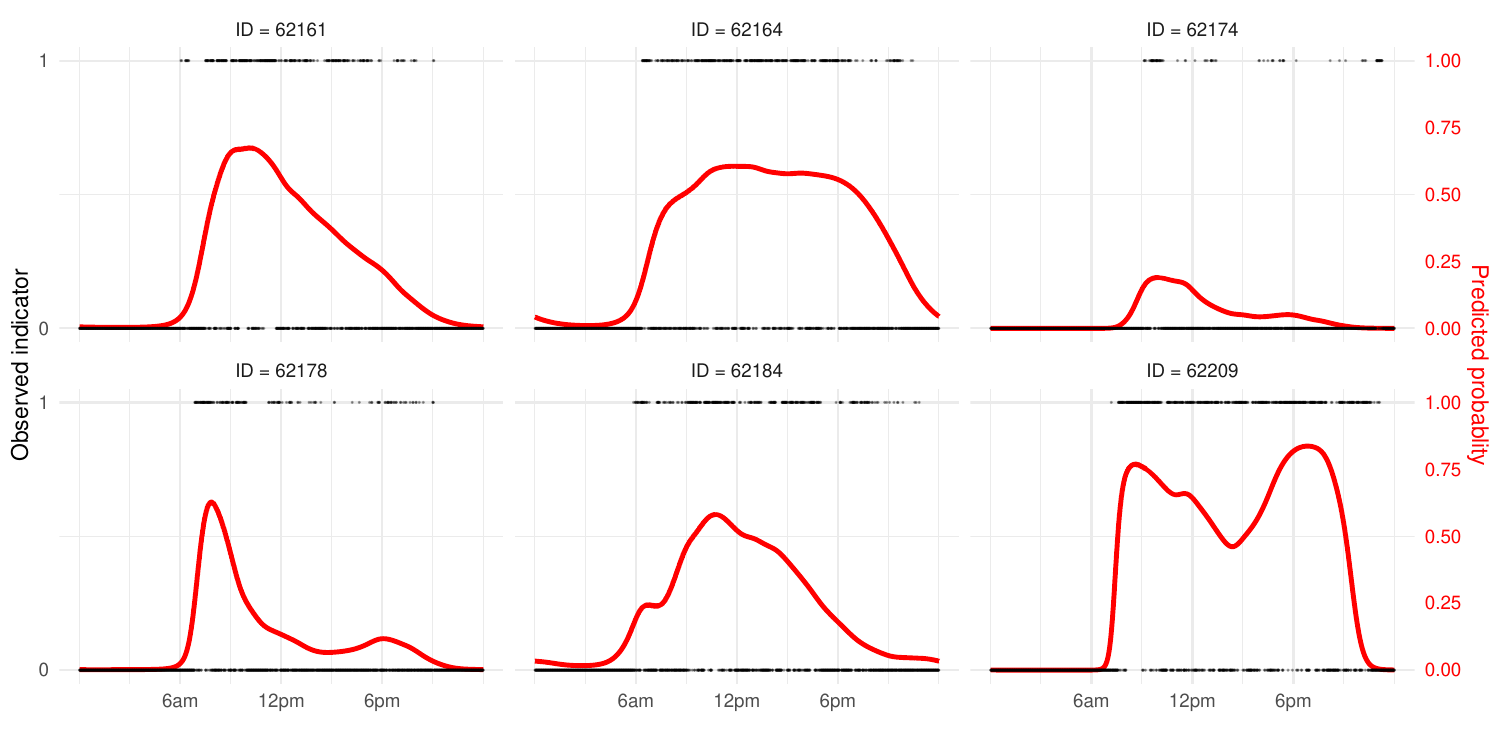}
\caption{Observed binary activity indicators and predicted probability curves of six participants over one day in the NHANES dataset.}   
\label{fig:nhanes_data}
\end{subfigure}

\begin{subfigure}{0.9\textwidth}
\includegraphics[width=\textwidth]{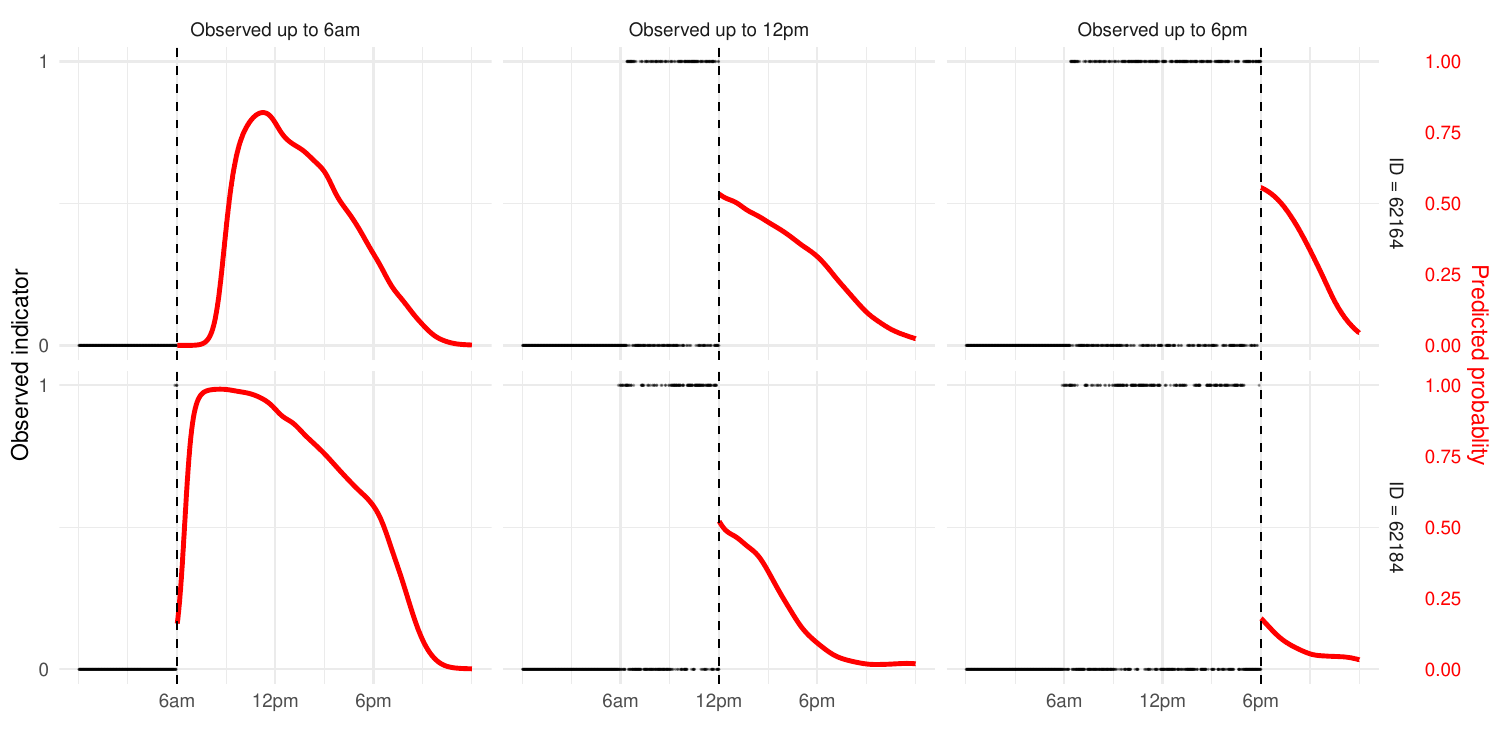}
\caption{Illustration of dynamic prediction of minute-level binary activity indicator using two participants from the NHANES study.}
\label{fig:nhanes_dynpred}   
\end{subfigure}

\caption{Visual illustration of the structure of the NHANES minute-level activity indicator as a motivation example for the dynamic prediction problem. The black dots indicate the observed activity status and the red lines represent the predicted probability of an active minute based on given observations.}
\label{fig:MotiveExp}
\end{figure}

Our goal is to predict future (later in the day) probability that an individual is active given their historical (earlier in the day) data. Given the minute-level activity indicators during a certain time interval, we hope to make personalized predictions of the afterward activity pattern that are accurate, efficient and can update as the observed interval extends. Figure \ref{fig:nhanes_dynpred} is an illustration of this problem using the activity indicator tracks of two subjects. The black dots are observed binary indicators, taking on values of either 0 (inactive) or 1 (active), and the dashed vertical lines represent the end of observation. From left to right, the three panels represent the increasing length of the observed interval, respectively 0-6am, 0am-12pm, and 0am-6pm. With these observations earlier in the day, the minute-level probability of activity was predicted and visualized with the red curve, which changes as more observations are collected.

\section{Method}
\label{sec:method}

\subsection{GFPCA}
\label{subsec:gfpca}

We first introduce assumptions and notation that will be used throughout this paper. The data we observe is a series of generalized outcomes collected densely across a functional domain (e.g., time/space), whose distribution can be specified by a continuous latent function. Specifically, let $i=1,\ldots,N$ denote subject. The outcome of interest is a realization of a non-Gaussian random variable ($Y_i(t)$) observed discretely along a domain $t \in T$, generated by a continuous latent function ($\eta_i(t)$). We assume that the outcome follows an exponential family distribution through a link function $g$: $g[E(Y_i(t))] = \eta_i(t)$. Though functional data like this can be observed either densely or sparsely. Here, like our motivating example in Section \ref{sec:data}, we assume that the data are collected densely along a common grid for all participants, ${t_j, j = 1..J}$, along the functional domain $T$. In the NHANES data, $t_j$ is each minute of the day, and $Y_i(t_j)$ is the binary active/inactive indicator associated with the $t_j$ minute. 

The GFPCA framework assumes that for each individual $i$, their latent function consists of a population-level functional fixed effect and an individual-specific functional random effect: $g(Y_i(t)) = \eta_i(t) = f_0(t)+ b_i(t), \hspace{0.2cm} t  \in T$. $b_i(t)$, the subject-specific random effect, is a zero-mean Gaussian process characterized by an unknown covariance operator $\Sigma_b$: $\Sigma_b(u, t) = Var(b_i(u), b_i(t))$, $u,t \in T$.

If $b_i$ is highly nonlinear, it can be modeled using standard GLMM software by specifying a sufficiently flexible spline basis for the random effect, $b_i(t) = \sum_{k=1}^K \xi_{ik} \phi_k(t)$ with $\phi_k$ being the basis functions, $K$ indicating the total number of basis functions used for estimation, and $\boldsymbol{\xi}_{i} \sim N(0, \Sigma_\xi)$. However, for most spline bases, the specification of the covariance matrix for the subject-specific coefficients ($\Sigma_\xi$) is not generally known in advance, and an unstructured covariance matrix is most likely appropriate. For highly nonlinear $b_i$, $K$ may be relatively large, resulting in a total of $K \times (K-1)/2$ variance parameters to be estimated. As a result, model estimation is often infeasible using traditional methods for generalized linear mixed models due to the heavy computational burden associated with a large number of variance parameters, especially with a large sample size and high density of observed data (e.g., $N = 4561$ and $J = 1440$ in the NHANES data). Moreover, when the intent is to obtain out-of-sample predictions, the individual random effects of new subjects can not be directly predicted using most existing software without re-estimating part or all of the parameters, which imposes further computational burden. 

As an alternative to general purpose spline bases (e.g., B-splines, cubic regresison lines, etc.), one may consider a more parsimonous model which is the basis for GFPCA methodology. By the Karhunen-Loeve theorem, $b_i$ can be represented as an infinite linear combination of orthogonal functions. The GFPCA model then becomes: $
    g(Y_i(t)) = \eta_i(t) = f_0(t)+\sum_{k=1}^{\infty}\xi_{ik}\phi_k(t), \hspace{0.2cm}
    t  \in T$, where the basis functions $\phi_k(t)$ are pairwise orthogonal and the coefficients $\xi_{k}$ are mutually independent random variables. If the function $\eta_i(t)$ is Guassian and the $\phi_k(t)$ are orthonormal, then the subject-specific coefficients $\xi_k$ are independent zero-mean Gaussian random variables:  $\xi_{ik} \sim N(0, \lambda_k)$ with decreasing variance such that $\lambda_1 > \lambda_2 >...$. This expansion is also known as the eigenvector transformation, with $\phi_k(t)$ as 
 eigenfunctions and $\lambda_k$ as eigenvalues. In practice this infinite sum is approximated using a relatively small number of basis functions as higher order bases often contribute relatively little to the percentage of variance explained. The model then becomes:
\begin{equation}
   \eta_i(t) = f_0(t)+\sum_{k=1}^{K}\xi_{ik}\phi_k(t), \hspace{0.3cm}
    t  \in T.
    \label{eq:gfpca}
\end{equation}

The latent function in Model \ref{eq:gfpca} consists of a
\textit{population} mean function $f_0(t)$ and an \textit{individual-specific} process $\sum_{k=1}^{K}\xi_{ik}\phi_k(t)$, the latter is a linear combination of $K$ functional principal components (FPC) $\phi_k(t)$. The FPCs are shared across population, but the coefficients $\xi_{ik}$ is specific to each subject. If we make an analogy to GLMM, then $f_0(t)$ can be perceived as a functional ``fixed effect", and $\xi_{ik}$ are similar to the ``random slopes" in the random effect. Because the $\xi_{ik}$ are independent within subjects, the number of variance parameters to be estimated has now been reduced from $K \times (K-1)/2$ to $K$, a substantial reduction in computational complexity. However, the primary methodologic challenge is how to estimate the eigenfunctions.

Many methods of parameter estimation for Model \ref{eq:gfpca}  have been proposed. However, there is a lack of straightforward implementation in practice, especially for large datasets with complex structures. Many existing methods are limited due to computational efficiency and/or model flexibility. For example, \citet{hall2018} modelled the generalized outcomes using a latent Gaussian process model, with experimentation on small, sparse dataset. The method assumes variation of the latent process about its population mean is relatively small. 
\citet{gertheiss2017} proposed and compared a two-stage estimation approach and a Bayesian joint estimation approach, with a simulation on sparse repeated measures generated from a latent process composed of two eigenfunctions. The generation mechanism is limited in both scale and flexibility. \citet{Li2018} proposed an efficient algorithm for FPCA on exponential data, under the assumption that a low-rank Singular Value Decomposition (SVD) exists of the natural parameters. The density of observation is greater in this paper compared to the previous ones, with 50 repeated measures from each subject. However, it is still far below the NHANES activity indicator. Its performance and efficiency on large, dense datasets remains to be explored. In terms of implementation, the software or package development of these proposed methods has been slow. Examples include a variational EM algorithm proposed by \citet{wrobel2019} for binary functional data, accompanied by the \textit{registr} package on GitHub \citep{Wrobel2018}.

\subsection{fGFPCA}
\label{subset:fGFPCA}

Despite the lack of efficient estimation or straightforward implementation of fGFPCA on generalized functional data, for Gaussian functions there exists fast implementation of esimtation methods, such as the Fast Covariance Estimation (FACE) algorithm \citep{face}. \citet{fGFPCA} combines the computational efficiencies of FACE with data binning to create a fast implementation of GFPCA (fGFPCA). The method speeds up the computation significantly by estimating the latent function $\eta_i$ on a binned grid slightly sparser than the observation grid, followed by an application of fast methods for FPCA (e.g., FACE) to the binned data to smooth across bins. A final normalization step follows to obtain estimates on the non-binned domain.

The fGFPCA algorithm is laid out as the following steps: 

\begin{enumerate}
    \item Bin observations

Here we bin the densely collected data into small, non-overlapping intervals with equal length $w$. We denote the index of bins as $s = 1,...,S$, and the index of the midpoint of the sth bin as  $m_s$. The sth bin includes observations collected over $(t_{m_s}-\frac{w}{2}, t_{m_s}+\frac{w}{2}]$.

The binning step assumes neighboring observations are generated from the same value of latent function $\eta_i(t_{m_s})$, thus allowing us to pool information from neighboring observations and estimate $\eta_i(t_{m_s})$ at the expense of minor misspecification on a slightly sparser grid. we Note that this paper set up bins as equally-spaced for simplicity. However, this method can be applied to unequally spaced bins as well. 
    
    \item Fit local intercept-only generalized mixed models within each bin

To estimate the latent function $\eta_i(t_{m_s})$, we fit a local Generalized Linear Mixed Model (GLMM) within every bin. These models include only a fixed and a random intercept, that is, at the $s^{\text{th}}$ bin: 
\begin{equation}
    g[E(Y_i(t_j))] =\eta_i(t_{m_s})= \beta_0(t_{m_s})+b_i(t_{m_s}),
    \hspace{0.5cm}t_j \in (t_{m_s}-\frac{w}{2}, t_{m_s}+\frac{w}{2}].
    \label{eq:local_glmm}
\end{equation}

With Model \ref{eq:local_glmm} above, an individual-specific intercept $\hat{\eta}_i(t_{m_s})$ can be estimated within each bin. Connecting these values from the same subject across all bins, we will get the estimated latent functions  $\hat{\eta}_i$ for each individual on the binned grid $S$.
    
    \item Implement FPCA

Since the estimated latent function $\hat{\eta}_i(t_{m_s})$ is continuous, we could apply the FPCA algorithm and fit the following model: 
\begin{equation}
    \hat{\eta}_i(t_{m_s}) = f_0(t_{m_s})+\sum_{k=1}^K\xi_{ik}\phi_{k}(t_{m_s})+\epsilon_i(t_{m_s}).
    \label{eq:fpca}
\end{equation}

The FPCA model produces estimates of smooth FPCs $\hat{\phi}_k(t_{m_s})$. The
number of PC functions, $K$, is often determined by the cumulative percentage of variance explained. \citet{zhou2024} also proposed selection using sum of squared 2nd-order differences (SSSOD).

    \item Re-evaluate parameters using a global GLMM

As mentioned in step 1, the mis-specification of local models \ref{eq:local_glmm} will lead to biased estimates, including variance of coefficients $\hat{\lambda}_k$ and population mean $\hat{f}_0$. This step corrects this bias by fitting a global GLMM, re-estimating $f_0$ as the fixed-effect and $\lambda_k$ as the random-effect. We use the term ``global" as it is uses all the observations on the original observation grid, unlike the ``local" models in Step 2 which use only obervations within a bin. 

Note that the PC functions $\hat{\phi}_k$ from Step 3 are estimated on the binned grid. To fit the global model on the observations grid, we need to extend the $\hat{\phi}_k(t_{m_s})$ back to the observation grid $\hat{\phi}_k(t_j)$. Many methods are applicable here since the estimated PC functions are not biased. In this paper, we project grid S back to grid J using regression on a rich set of spline basis functions: $
g[E(Y_i(t_j))|\hat{\phi}_k(t_j)] = f_0(t_j)+\sum_{k=1}^K \xi_{ik}\hat{\phi}_k(t_j)
$.

From this model, parameter estimates that will be used later for dynamic prediction are obtained, including the population mean $\hat{f}_0(t)$ and variance of coefficients $\hat{\lambda}_k$. This model could be efficiently fit using the \textit{bam} function in the \textit{mgcv} packages in R. 
\end{enumerate}

In addition to fGFPCA, alternative methods exist to obtain the parameter estimates to be used for dynamic prediction. One potential choice was recently proposed by \citet{weishampel2023}. In this paper, each binary function is ``smoothed" separately using a general additive model (GAM), and FPCA was then fit on the collection of smoothed tracks. This method can be efficient in some scenarios, but does not work well for our motivation example of the physical activity data. Since a large proportion of subjects are sedatary at night, the hours of consecutive zeros can cause identifiability issues when fitting the individual GAM models. This shortcoming is illustrated in \citet{leroux2022} through a detailed comparison between this method and fGFPCA on the NHANES data. Another option is the \textit{registr} package by \citet{Wrobel2018} as we mentioned in the Introduction. It can achieve similar efficiency as fGFPCA \cite{fGFPCA}, but its application is restricted to binary data and cannot be directly applied to other exponential family distributions.

\subsection{Dynamic prediction}
\label{subsec:dyn_pred}

In this section we show how to use the results obtained from fGFPCA to obtain dynamic predictions. We formulate the problem in terms of partitioning the functional data along the domain in two parts and using data from one element of the partition to predict a participant's latent function in the other element of the partition. This corresponds to using historical data to predict the future if the functional domain is time. Note that this is our framework, but methods are generally applicable to any combination of observed data partition where data are either observed or unobserved.

Assume we have a subject, $u$, which was not used for model fitting. For this subject, we only observe the outcome along a portion of the domain: $Y_u(t_j)$, $j=1,...,J_u$ and $J_u < J$. With the population level estimates obtained from  Step 4 described in Section \ref{subset:fGFPCA}, including population mean $\hat{f}_0$, FPCs $\hat{\phi}_k$ and variance of individual scores $\hat{\lambda}_k$, the prediction at any point $t_j$ beyond the observed track ($j > J_u$) gets down to the estimation of out-of-sample coefficient $\hat{\xi}_{uk}$, as follows:

\[
\begin{aligned}
    \hat{\eta}_u(t_j)|\hat{\mathbf{\Theta}} &= \hat{f}_0(t_j)+\sum_{k=1}^K \hat{\xi}_{uk}\hat{\phi}_k(t_j)
    \hspace{0.3cm}  J_u < j < J  \\
    \hat{\mathbf{\Theta}} & = \{\hat{f}_0(t), \hat{\phi}_k(t), \hat{\lambda}_k, k = 1...K \}.
\end{aligned}
\]

Since the outcome $Y_u(t_j)$ follows an exponential family distribution characterized by the latent process $\eta_u(t_j)$, it is straightforward to write out the posterior log-likelihood of the observed track: 

\[
\begin{aligned}
    l_u = l(\mathbf{Y}_u | \hat{\boldsymbol{\xi}}_u, 
    \hat{\mathbf{\Theta}}) &=\sum_{j<J_u}log(h(Y_u(t_j)))+\eta_u(t_j)T(Y_u(t_j))-log(A[\eta_u(t_j)])\\
    \eta_u(t_j) & = \hat{f}_0(t_j)+\sum_{k=1}^K\xi_{uk}\hat{\phi}_k(t_j)\\
    \mathbf{Y}_u &= \{Y_u(t_j), j = 1,...,J_u \}, \hspace{0.3cm} \hat{\boldsymbol{\xi}}_u = \{\hat{\xi}_{uk}, l = 1,..., K \}\\
\end{aligned}
\]

The form of $h(.)$, $T(.)$, and $A(.)$ depends on the specific type of exponential family distribution. We also know the prior distribution of coefficients: $\xi_{uk} \sim N(0, \hat{\lambda}_k)$. Therefore, the maximum likelihood estimator of the subject-specific coefficients can be obtained as:

\[
    \hat{\boldsymbol{\xi}}_u = 
    \underset{\boldsymbol{\xi}_u}{\mathrm{argmin}} \ L(\boldsymbol{\xi}_u|\boldsymbol{Y}_u, 
    \hat{\mathbf{\Theta}}) = \frac{p(\boldsymbol{Y}_u|\boldsymbol{\xi}_u, \hat{\mathbf{\Theta}})p(\boldsymbol{\xi}_u|\hat{\mathbf{\Theta}})}{\int p(\boldsymbol{Y}_u|\boldsymbol{\xi}_u, \hat{\mathbf{\Theta}})p(\boldsymbol{\xi}_u|\hat{\mathbf{\Theta}})d\boldsymbol{\xi}_u}
\]
With generalized outcomes, the likelihood above often does not have a closed-form solution. In such cases, we could either apply numeric maximization or use methods of approximation. For example, in R, the likelihood can be optimized using MCMC sampling and the L-BFGS (limited Broyden–Fletcher–Goldfarb–Shanno) algorithm from the \textit{rstan}. Alternative methods, such as Laplace Approximation, are also available though we found them to be numerically unstable in our data application and thus do not explore them further here.

The construction of prediction intervals is also straightforward according to likelihood theory. The covariance matrix of MLE is asymptotically the inverse of observed information matrix. We are able to get the pointwise posterior variance estimates of the latent function by taking the second derivative of the posterior likelihood conditioning on $\hat{f}_0(t_j)$, $\hat{\phi}_k(t_j)$ and $\hat{\lambda}_k$, as follows:

\[
Var(\hat{\eta}_u(t_j)|\hat{\mathbf{\Theta}})  = Var(\hat{f}_0(t_j)+\sum_{k=1}^K\hat{\phi}_k(t_j)\hat{\xi}_{uk})
=\sum_{k=1}^K \hat{\phi}_k^2(t_j)\hat{\lambda}_k,
\hspace{0.3cm}  J_u < j < J.
\]

Wald prediction interval could be constructed with this variance estimator at each point along the functional domain. In practice, when using Bayesian methods for parameter estimation, we could also construct a ``credible" prediction interval by taking the sampling quantiles. Given $\hat{\mathbf{\Theta}}$, the two intervals are very similar numerically in our simulations and data application, though this may not be true in general.

\section{Simulation study}
\label{sec:simulation}

\subsection{Data generation scheme}
\label{subser:gen_data}

In this section, we implement a simulation study to illustrate the predictive performance and
computational efficiency of the proposed method. The functional outcome is generated from a cyclic latent process as follows:

\[
\begin{aligned}
Y_i(t) & \sim Binomial(\frac{exp(\eta_i(t))}{1+exp(\eta_i(t))}), \hspace{0.3cm} t  \in [0, 1] \\
\eta_i(t) &=\xi_{i1}\sqrt{2}sin(2\pi t)+\xi_{i2}\sqrt{2}cos(2\pi t)+\xi_{i3}\sqrt{2}sin(4\pi t)+\xi_{i4}\sqrt{2}cos(4\pi t)\\
\end{aligned}
\]

The latent process has a zero mean for simplicity ($f_0(.) = 0$). The true FPCs $\phi_k(t) = \{\sqrt{2}sin(2\pi t), \sqrt{2}cos(2\pi t), \sqrt{2}sin(4\pi t), \sqrt{2}cos(4\pi t) \}$. Coefficients are generated independently from zero-mean Gaussian distribution with decreasing variance: $\xi_{ik} \sim N(0, 0.5^{k-1})$, $k=1, 2, 3, 4$. For each subject, we generate 1000 observations ($J = 1000$) along $t \in [0, 1]$ on a regular grid. Every simulated dataset is split into the ``training set" used for model estimation, and a ``test set" for evaluating out-of-sample model performance. That is, the model estimated using only the training set is used to predict future observations of subjects in the test set based on their partially observed tracks. The proposed method is compared to existing methods that are commonly used for similar purposes. However, the density of the simulated data is high and beyond the capacity of most these methods (see Section \ref{subsec:compete_methods} for details). Therefore we set up two scenarios with different training sample size: 1) a larger training set of 500 subjects ($N_{train} = 500$); and 2) a smaller training set of 100 subjects ($N_{train} = 100$). The test sample sizes of the two scenarios are the same ($N_{test} = 100$). In the second scenario, we also need to reduce the measurement density, taking only one every ten observations and discarding the rest.

\subsection{Competing Approaches}
\label{subsec:compete_methods}

In addition to fGFPCA, we consider two competing methods as a basis for comparison.

\subsubsection{GLMMadaptive}
\label{subsubsec:adglmm}

``GLMMadaptive" is the abbreviation of ``Generalized Linear Mixed Models using Adaptive Gaussian Quadrature". Developed by \citet{Rizopoulos2017}, this method fits a generalized mixed effect model and obtains estimates by approximating the integral over the random effects using the adaptive Gauss-Hermite quadrature rule, accompanied by an R package \textit{GLMMadaptive}. Compared to traditional methods in $\it{lme4}$ and $\it{nlme}$, the approximation has improved the efficiency substantially and offers functionality for making dynamic predictions. According to \citet{Stringer2023}, \textit{GLMMadaptive} can fit the same model 2-4 times faster than \textit{lme4} without compromising performance. However, with the measurement density of our simulated data, this method is still not ``fast" enough when using models with complex, non-linear subject-specific effects. That is, GLMMadaptive is only feasible (within reasonable time) at the cost of model flexibility. With this restriction, GLMMadaptive is only able to estimate a random intercept and slope model on the ``larger" training dataset ($N_{train}=500, J=1000$). Specifically, the model:
\begin{equation}
g(E(Y_i(t_j))) = \beta_0+\beta_1t_j+b_{i0}+b_{i1}t_j.   
    \label{eq:glmm_linear}
\end{equation}

This model essentially has a linear structure and is clearly not sufficiently flexible to adequately model the complex data generation mechanism. Therefore, we would expect low predictive accuracy compared to the proposed method.

In the second scenario, the smaller training size makes it possible to increase the flexibility of the GLMMadaptive model by incorporating a small set of spline basis functions: 

\begin{equation}
    g(E(Y_i(t_j))) = \sum_{k=1}^4\zeta_{k}B_k(t_j)+\sum_{l=1}^4\xi_{il}\phi_l(t_j),
    \label{eq:glmm_sp}
\end{equation}

where the $\phi_l(.)$ are natural cubic splines.

However, to reduce the model fitting time within a reasonable range, we still need to reduce the observations density by taking only one out of ten neighboring observations. The means 90\% of the observations are discared the the informaiton within cannot be utilized. 

\subsubsection{GFOSR}
\label{subsubsec:gfosr}

Generalized Function on Scalar Regression (GFOSR) estimates the association between a functional response and scalar covariates {\color{purple}\citep{Reiss2010, fda2005, Bauer2018}}. In our context, this framework allows for directly modeling the future trajectories (functional outcome) explicitly conditional on the scalar covariates (historical/observed data). Specifically, we predict future trajectories using several of the most recent observations from the observed track. The concept is similar to a distributed lag model approach \citep{Baek2022}, though the coefficients are functional and not necessarily decaying. For a new subject $u$, if the observed track stopped at $t_m$ ($t_m < T$, $m < J$), and we use the last $L$ observations as predictors, then the model used for predicting the future can be rigorously expressed as: 
\begin{equation}
g(E[Y_i(t_j)|Y_i(v), v\leq t_m]) = \beta_0(t_j) +\sum_{l} \beta_{m-l+1} (t_j)Y_i(t_l), \hspace{0.3cm}
l  = m,...,m-(L-1),  \hspace{0.3cm}
t_j > t_m 
\label{eq:gfosr}  
\end{equation}

The larger L is, the more flexible this model can be, though without penalization, multicollinearity can be an issue. Here we limit our investigation to two different values: $L = 1$ or $L = 5$.

In our case, since the predictor (historical data) is itself non-Gaussian functional data, we could also consider a Function-on-Function regression framework \citep{Morris2015, fdaR2024,Goldsmith2011}. The FOSR and FOFR frameworks are both quite flexible, but for the latter, we are not aware of established methods for prediction where both the predictor and outcome are generalized functions. Here, we choose to use the FOSR approach. 

\subsection{Evaluation metrics}

Out-of-sample predictive performance is evaluated with two metrics: 1) Integrated Squared Error (ISE) and
Area Under the Receiver Operating Curve (AUC). ISE is the squared error of the predicted latent function over an interval of t: $ISE_i = \sum_{t_j}(\hat{\eta}_i(t_j)-\eta_i(t_j))^2$. It assesses the
prediction accuracy of a continuous latent process. 
The second metric, AUC, focuses on comparing the predicted latent function to the observed binary outcome. Both metrics are reported as average across subjects and simulated datasets. 

The uncertainty of the predictions are evaluated by prediction intervals and presented in the Supplement. For the proposed method fGFPCA, we calculated the credible interval using the sampling quantiles of latent function at each measurement point. This interval directly expresses the probability that the true value of $\eta$ lies within the range. For the completing methods, GFOSR and GLMMadaptive, Wald intervals are calculated using the standard error estimates obtained directly from the corresponding packages.

\subsection{Results}

\subsubsection{Large training set}
\label{sss:sim_result_large}

Here we fit the fGFPCA model on a larger training set with 500 subjects and compare its predictive performance on an additional 100 subjects to the simple linear GLMMadaptive model \ref{eq:glmm_linear} and the GFOSR model \ref{eq:gfosr} with $L=1$ and $L=5$. In the test set, future predictions are made based on observations given up to certain cut-off points $t_{J_u} \in \{0.2, 0.4, 0.6, 0.8\}$. That is, given observations along 0-0.2; prediction is made on 0.2-1; given observations along 0-0.4, prediction is made on 0.4-1, and so forth. The prediction performance is presented by equal length intervals: (0.2, 0.4], (0.4, 0.6], (0.6. 0.8] and (0.8, 1], so that the predictive performance is comparable on an equal number of observations and across common grids. 

Figure \ref{fig:sim1} presents an example of predictions for two randomly selected test subjects from one of the 500 simulated datasets. As the figure reveals, the four methods provide predictions with very different degrees of nonlinearity. fGFPCA results in the most flexible predicted tracks and is largely capable of capturing the individual and cyclic patterns of latent functions. The shape of GFOSR prediction depends on the number of observations used. With the last five observations, the predicted curves resemble quadratic or cubic functions without periodicity. When using only the last observation, the predictions hardly reveal any individual-specific deviation from the population. This can be explained by the model formulation of GFOSR in Equation \ref{eq:gfosr}. Since this model does not include random effects and there are only two possible outcomes for each observation (0 or 1), there would be only two unique predicted tracks across the test sample when $L=1$. Finally, the GLMMadaptive model results in monotone predictions due to the linear random effects structure on the log-odds scale. We reiterate that this is a computational limitation of the current implementation in our simulation scenario and not a limitation of the method in general. 

ISE and AUC are summarized in Table \ref{tab:sim_large}, with lower values of ISE and higher values of AUC representing better predictive performance. fGFPCA provided the best predictive performance in all scenarios and won over the other three methods by a large margin. The improvement of ISE across all scenarios ranges from 13.1\% to 97.3\%, with an average improvement of 63.9\%. AUC improved from 4.5\% to 48.1\%, with an average of 24.7\%. Ideally for dynamic prediction, the accumulation of additional data would result in improved performance. As the maximum observation time extends from 0.2 to 0.8, the predicted tracks from the proposed method move closer to the true latent functions. The other methods, on the other hand, do not change much as observation tracks extend. The more flexible fGFPCA on the left side in Figure \ref{fig:sim1} clearly shows this behavior. Table \ref{tab:sim_large} shows this result holds across subjects and simulated datasets. When comparing the same prediction window across different maximum observation times (rows), ISE decreases significantly, and AUC increases for fGFPCA as the observation track gets longer. However, the other three methods did not show such improvement. GLMMadaptive actually resulted in worse performance with more observations.

Regarding computation time, GFOSR is the fastest among the four models, taking about 1.5 seconds for each simulated dataset when using the last observation and 9 seconds when using the last five observations. However, the low accuracy of the predictions limits the appeal of very fast computation in this context. In contrast, fGFPCA takes approximately 5.8 minutes for each simulated dataset, while the linear GLMMadaptive runs slightly faster at 3.35 minutes on average.

\subsubsection{Small training set}
\label{sssec:sim_result_small}

In this section, we present the results when the training sample size and measurement density are reduced to accommodate a more flexible GLMMadaptive estimated with a nonlinear subject-specific random effect using a low rank natural spline basis. The model Equation \ref{eq:glmm_sp} still takes a very long time to fit on the full gird even with smaller training set. Therefore, as mentioned in Section \ref{subser:gen_data}, we reduced the data density by taking one out of every 10 neighboring observations, causing 90\% of the observations discarded. The out-of-sample prediction, on the other hand, is on the original measurement grid. The other three methods (fGFPCA and GFOSR with the last one or five observations) did not need data reduction and utilized all information in the training samples. 

We also note that the GLMMadaptive model is not as numerically stable as the other three methods since consecutive zeros or ones can cause extreme values of parameter estimation. As a result, the mixed model occasionally runs into convergence issues. The simulated datasets with numeric problems are discarded. In the summary of predictive performance in Table \ref{tab:sim_small}, the AUC and ISE of GLMMadaptive are averaged over 496 datasets without numeric problems. The other three methods did not suffer from such issues and thus averaged over all 500 datasets. 

\begin{figure}
    \centering

    \begin{subfigure}{\textwidth}
    \includegraphics[width=\textwidth]{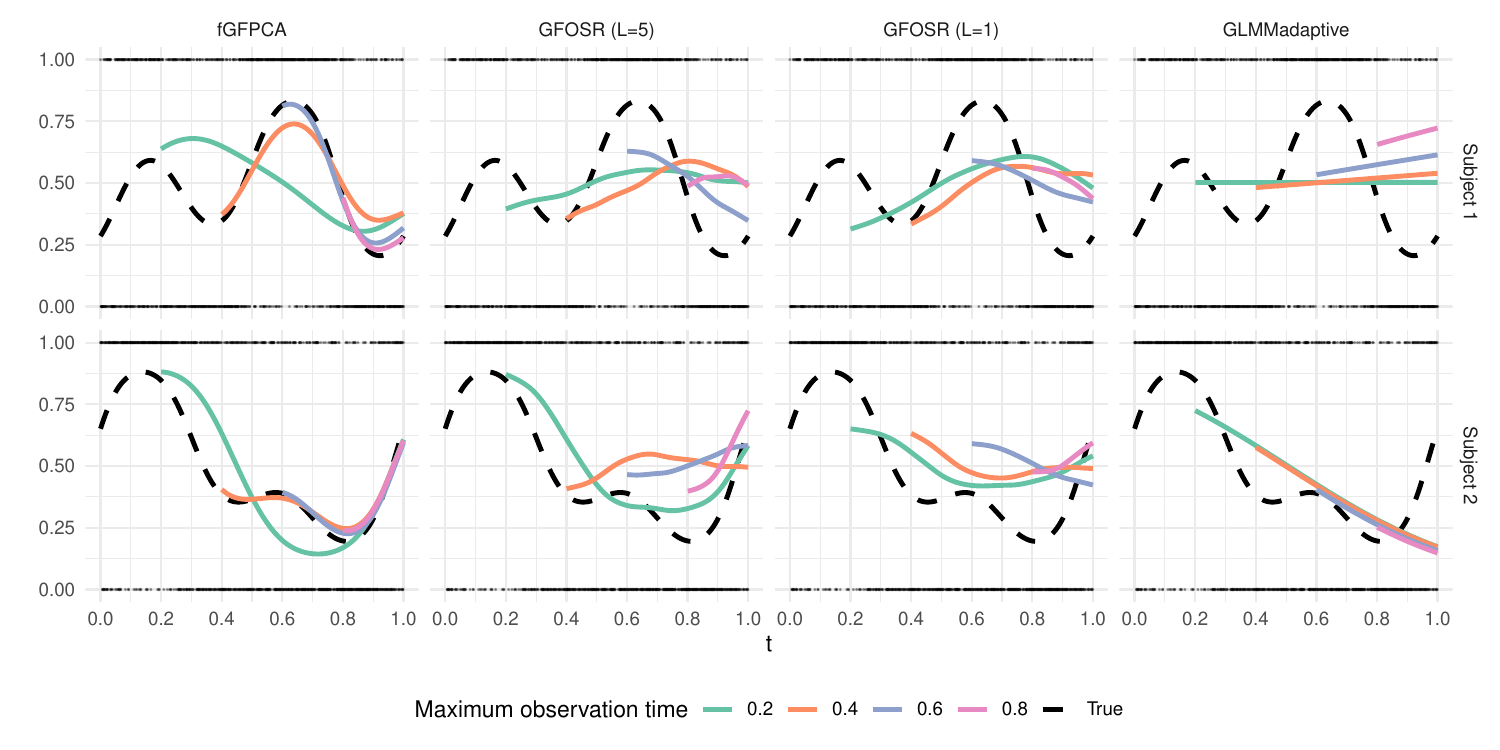}
    \caption{Training sample size = 500}
    \label{fig:sim1}
    \end{subfigure}

    \begin{subfigure}{\textwidth}
    \includegraphics[width=\textwidth]{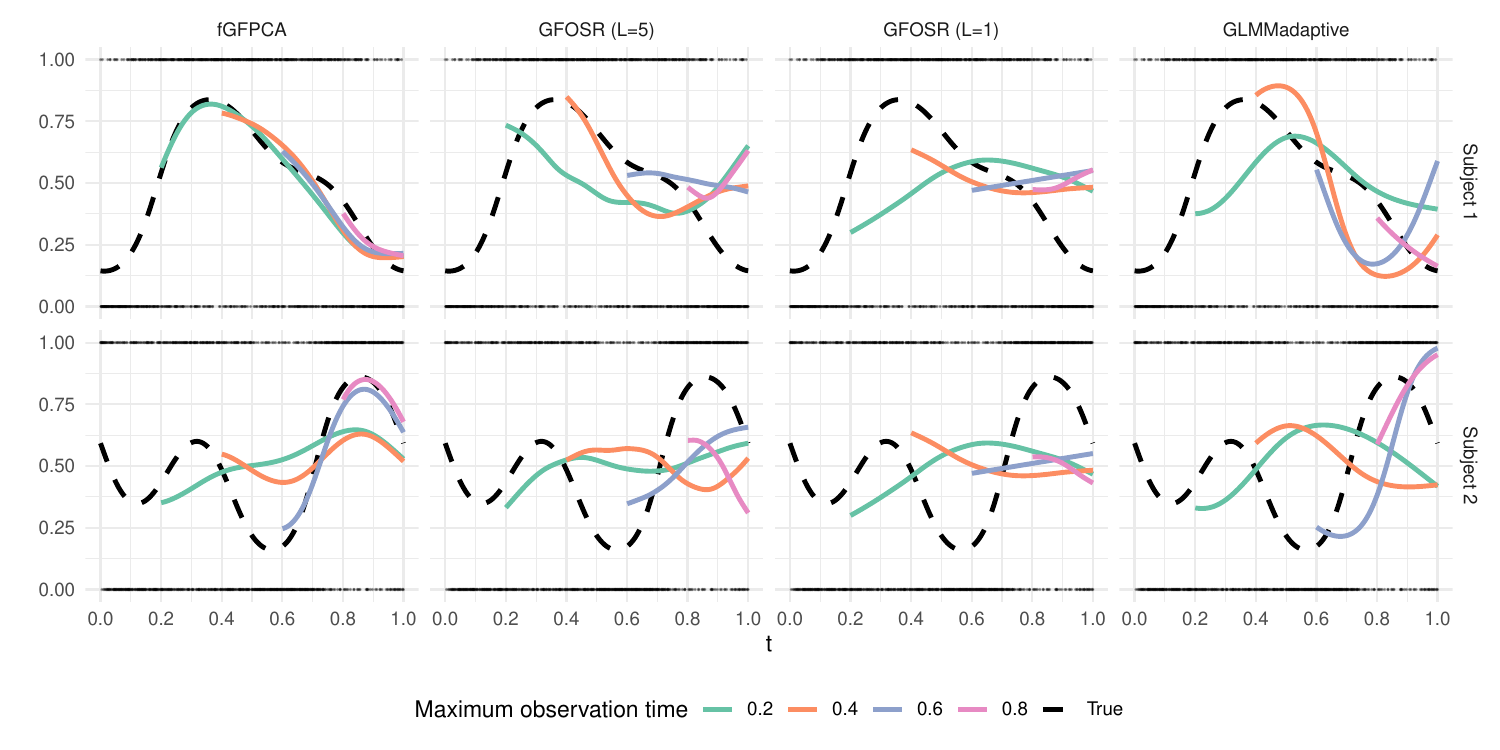}
    \caption{Training sample size = 100}
    \label{fig:sim2}
    \end{subfigure}
    
    \caption{Individual out-of-sample dynamic prediction of the latent process of two randomly selected subjects from the simulated data . The top panel shows prediction from model fit on 500 subjects, and the bottom panel 100 subjects. The black dots indicate observed outcome and black dashed line the true latent function. Solid colored lines are the predicted track, with the color of lines indicating different maximum observation time. }
\end{figure}

Figure \ref{fig:sim2} presents predictions for two participants in the same format as~\ref{fig:sim1}. As a result of increased model flexibility for GLMMadaptive, individual prediction tracks on the right column in Figure \ref{fig:sim2} are much more nonlinear when compared to the previous section, and also approach the true latent function as observed tracks are updated. However, predictions from this model are still considerably worse than fGFPCA, as seen in Table \ref{tab:sim_small}, with fGFPCA resulting in 11.1\% - 94.6\% (average 61.5\%) improvement in ISE and 0.9\% - 45.1\% (average 19.5\%) improvement in AUC across all scenarios. Computationally, since the size of the training data was reduced, both fGFPCA and GFOSR spent less time on model fitting. On average, fGFPCA took 5.23 minutes, while GFOSR took 0.6 and 4.4 seconds for $L=1$ and $L=5$  respectively. However, for GLMMadaptive, the increased flexibility offset computational gains associated with a decreased sample size, taking 6.35 minutes to fit on average. 

\begin{landscape}\begin{table}

\begin{subtable}{\textwidth}
\small
\centering
\begin{tabular}{lrrrrrrrrrrrrrrrr}
\toprule
\multicolumn{1}{c}{ } & \multicolumn{16}{c}{Maximum observation time} \\
\cmidrule(l{3pt}r{3pt}){2-17}
\multicolumn{1}{c}{ } & \multicolumn{4}{c}{fGFPCA} & \multicolumn{4}{c}{GFOSR (L=5)} & \multicolumn{4}{c}{GFOSR (L=1)} & \multicolumn{4}{c}{GLMMadaptive} \\
\cmidrule(l{3pt}r{3pt}){2-5} \cmidrule(l{3pt}r{3pt}){6-9} \cmidrule(l{3pt}r{3pt}){10-13} \cmidrule(l{3pt}r{3pt}){14-17}
Window & 0.2 & 0.4 & 0.6 & 0.8 & 0.2 & 0.4 & 0.6 & 0.8 & 0.2 & 0.4 & 0.6 & 0.8 & 0.2 & 0.4 & 0.6 & 0.8\\
\midrule
\addlinespace[0.3em]
\multicolumn{17}{l}{\textbf{ISE}}\\
\hspace{1em}(0.2, 0.4] & 184.19 &  &  &  & 274.95 &  &  &  & 362.48 &  &  &  & 387.71 &  &  & \\
\hspace{1em}(0.4, 0.6] & 218.57 & 113.45 &  &  & 277.44 & 220.21 &  &  & 286.61 & 262.55 &  &  & 291.58 & 269.80 &  & \\
\hspace{1em}(0.6, 0.8] & 274.57 & 76.51 & 27.07 &  & 322.34 & 373.11 & 325.02 &  & 385.70 & 410.51 & 389.31 &  & 315.78 & 282.74 & 278.24 & \\
\hspace{1em}(0.8, 1.0] & 119.59 & 106.93 & 25.46 & 16.41 & 290.98 & 318.08 & 350.85 & 333.58 & 328.48 & 341.27 & 354.21 & 347.07 & 563.01 & 477.49 & 597.75 & 600.34\\
\addlinespace[0.3em]
\multicolumn{17}{l}{\textbf{AUC}}\\
\hspace{1em}(0.2, 0.4] & 0.737 &  &  &  & 0.686 &  &  &  & 0.624 &  &  &  & 0.591 &  &  & \\
\hspace{1em}(0.4, 0.6] & 0.645 & 0.711 &  &  & 0.563 & 0.630 &  &  & 0.543 & 0.590 &  &  & 0.524 & 0.596 &  & \\
\hspace{1em}(0.6, 0.8] & 0.699 & 0.779 & 0.799 &  & 0.669 & 0.628 & 0.676 &  & 0.604 & 0.577 & 0.615 &  & 0.669 & 0.694 & 0.687 & \\
\hspace{1em}(0.8, 1.0] & 0.736 & 0.741 & 0.779 & 0.783 & 0.626 & 0.606 & 0.552 & 0.584 & 0.588 & 0.564 & 0.537 & 0.551 & 0.514 & 0.556 & 0.526 & 0.564\\
\bottomrule
\end{tabular}
\caption{Training sample size = 500}
\label{tab:sim_large}
\end{subtable}

\hfill

\begin{subtable}{\textwidth}
\small
\centering
\begin{tabular}{lrrrrrrrrrrrrrrrr}
\toprule
\multicolumn{1}{c}{ } & \multicolumn{16}{c}{Maximum observation time} \\
\cmidrule(l{3pt}r{3pt}){2-17}
\multicolumn{1}{c}{ } & \multicolumn{4}{c}{fGFPCA} & \multicolumn{4}{c}{GFOSR (L=5)} & \multicolumn{4}{c}{GFOSR (L=1)} & \multicolumn{4}{c}{GLMMadaptive} \\
\cmidrule(l{3pt}r{3pt}){2-5} \cmidrule(l{3pt}r{3pt}){6-9} \cmidrule(l{3pt}r{3pt}){10-13} \cmidrule(l{3pt}r{3pt}){14-17}
Window & 0.2 & 0.4 & 0.6 & 0.8 & 0.2 & 0.4 & 0.6 & 0.8 & 0.2 & 0.4 & 0.6 & 0.8 & 0.2 & 0.4 & 0.6 & 0.8\\
\midrule
\addlinespace[0.3em]
\multicolumn{17}{l}{\textbf{ISE}}\\
\hspace{1em}(0.2, 0.4] & 191.05 &  &  &  & 286.85 &  &  &  & 367.28 &  &  &  & 374.47 &  &  & \\
\hspace{1em}(0.4, 0.6] & 224.08 & 118.19 &  &  & 286.63 & 227.23 &  &  & 290.78 & 265.81 &  &  & 268.20 & 462.62 &  & \\
\hspace{1em}(0.6, 0.8] & 283.93 & 81.57 & 30.16 &  & 328.61 & 374.57 & 326.52 &  & 386.71 & 408.43 & 388.43 &  & 319.28 & 286.71 & 233.55 & \\
\hspace{1em}(0.8, 1.0] & 124.41 & 114.15 & 28.87 & 18.95 & 301.17 & 324.82 & 353.89 & 335.14 & 331.49 & 342.94 & 354.80 & 347.63 & 228.24 & 469.24 & 331.53 & 131.91\\
\addlinespace[0.3em]
\multicolumn{17}{l}{\textbf{AUC}}\\
\hspace{1em}(0.2, 0.4] & 0.734 &  &  &  & 0.679 &  &  &  & 0.620 &  &  &  & 0.672 &  &  & \\
\hspace{1em}(0.4, 0.6] & 0.644 & 0.710 &  &  & 0.549 & 0.627 &  &  & 0.536 & 0.590 &  &  & 0.622 & 0.671 &  & \\
\hspace{1em}(0.6, 0.8] & 0.695 & 0.775 & 0.800 &  & 0.656 & 0.617 & 0.668 &  & 0.598 & 0.574 & 0.613 &  & 0.689 & 0.698 & 0.730 & \\
\hspace{1em}(0.8, 1.0] & 0.735 & 0.739 & 0.778 & 0.782 & 0.611 & 0.588 & 0.551 & 0.578 & 0.580 & 0.556 & 0.536 & 0.551 & 0.680 & 0.629 & 0.678 & 0.743\\
\bottomrule
\end{tabular}

\caption{Training sample size = 100}
\label{tab:sim_small}
\end{subtable}

\caption{Predictive performance of models fit on simulated data. Both the Integrated Squared Error (ISE) and AUC are averaged across all simulated datasets without numeric issues.}
\label{tab:sim}
\end{table}
\end{landscape}

\section{Case study}
\label{sec:case_study}

This section explores the application of the proposed method to a real-world dataset. Specifically, we conducted a case study is on the NHANES 2011-2014 minute-level active/inactive indicators introduced in Section \ref{sec:data}. We aim to predict active/inactive states later in the day using active/inactive status observed earlier in the day. Similar to the simulation study, the sample was split into a training set ($N_{train}$ = 5257) and a test set ($N_{test}$ = 3506). The models are fit in the training set and then evaluated in the test set. Since we do not know the true latent functions in this case, predictive performance is evaluated using AUC alone. For these data, we were only able to fit the GLMMadaptive model as in \eqref{eq:glmm_linear}.

\begin{figure}
\centering
\includegraphics[width=\textwidth]{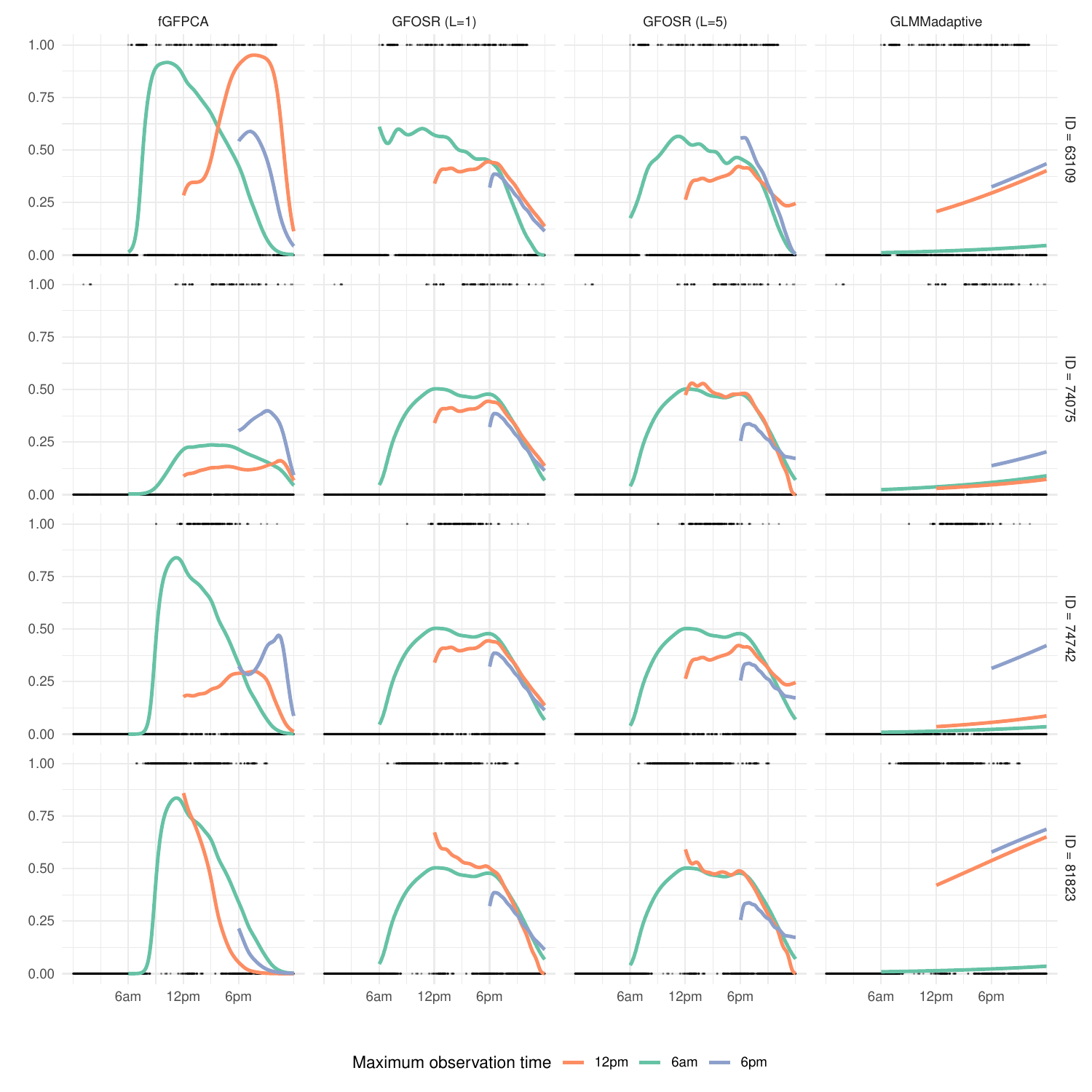}
\caption{Individual out-of-sample prediction tracks of the latent process of the NHANES data. The black dots indicate observed outcomes. Solid colored lines are the predicted track, with the color of lines indicating different maximum observation time. }
\label{fig:appl}
\end{figure}

Figure \ref{fig:appl} presents the individual prediction tracks of four randomly selected participants in the test set across different methods. The figure follows a similar format as Figures \ref{fig:sim1} and \ref{fig:sim2}, with red, green, and blue lines respectively indicating prediction obtained given observation from 0-6am, 0am-12pm and 0am-6pm. As expected, fGFPCA, shown in the left column, resulted in more nonlinearity in the predicted tracks and increased variability across individuals, neither of which is observed in predictions obtained from the three other methods. As in the simulation study, GLMMadaptive, shown in the right column, results in monotone predictions as it only includes a linear time effect. GFOSR provides more non-linear predictions but with substantially less heterogeneity between individuals when compared to fGFPCA. Additionally, the predictions from fGFPCA updates substantially as observations are accumulated, visually seen as the different shapes of tracks of different colors for the same prediction time. This behavior is less apparent in the other methods considered here, with mean shifts largely characterizing the difference in prediction tracks.

Table \ref{tab:appl} summarizes the AUC across different methods. This table is structured similarly to Tables \ref{tab:sim_large} and \ref{tab:sim_small}, where AUC is calculated on equal-length time windows for comparability (6am-12pm, 12pm-6pm, 6pm-12am). As the table shows, fGFPCA has the highest AUC in any situation, with the largest improvement seen for prediction windows further away from the observation window. For example, given data from 0am-6am, the AUC of fGFPCA is 0.118 higher than GLMM on 6am-12pm, but 0.192 higher on 6pm-12am. The improvement of predictive accuracy of fGFPCA is also more rapid as more observations are collected. In fact, the GFOSR and GLMM models showed worse performance as the prediction window moves away from the observation window, while fGFPCA declined and then improved. The latter finding is intuitively consistent with the fact that physical activity exhibits cyclic diurnal patterns. From these results, we conclude that fGFPCA is better capable of capturing and predicting the heterogeneous, individualized and nonlinear latent processes that characterize active/inactive patterns across the day.

\begin{table}
\small
\centering
\begin{tabular}{lrrrrrrrrrrrr}
\toprule
\multicolumn{1}{c}{ } & \multicolumn{12}{c}{Maximum observation time} \\
\cmidrule(l{3pt}r{3pt}){2-13}
\multicolumn{1}{c}{ } & \multicolumn{3}{c}{fGFPCA} & \multicolumn{3}{c}{GFOSR (L=5)} & \multicolumn{3}{c}{GFOSR (L=1)} & \multicolumn{3}{c}{GLMMadaptive} \\
\cmidrule(l{3pt}r{3pt}){2-4} \cmidrule(l{3pt}r{3pt}){5-7} \cmidrule(l{3pt}r{3pt}){8-10} \cmidrule(l{3pt}r{3pt}){11-13}
Window & 6am & 12pm & 6pm & 6am & 12pm & 6pm & 6am & 12pm & 6pm & 6am & 12pm & 6pm\\
\midrule
6am-12pm & 0.699 &  &  & 0.689 &  &  & 0.684 &  &  & 0.581 &  & \\
12pm-6pm & 0.534 & 0.711 &  & 0.519 & 0.699 &  & 0.520 & 0.654 &  & 0.532 & 0.701 & \\
6pm-12am & 0.706 & 0.689 & 0.774 & 0.677 & 0.678 & 0.723 & 0.677 & 0.676 & 0.708 & 0.514 & 0.565 & 0.626\\
\bottomrule
\end{tabular}
\caption{Predictive performance of models fit on the NHANES data}
\label{tab:appl}
\end{table}

\section{Discussion}
\label{sec:discussion}

In this paper, we proposed a fast, scalable method for dynamically predicting generalized functional data. Compared with existing methods, using generalized linear mixed model software, the proposed method offers substantial computational efficiencies and thus can be applied to larger samples with high density functional data. The prediction it makes is also much more dynamic, personalized and accurate, thus achieves the essential goals of dynamic prediction. The simulation study and data application presented in Sections \ref{sec:simulation} and \ref{sec:case_study}, respectively, illustrate the proposed approach's feasibility and improved performance in comparison to other methods when data exhibit highly non-linear latent trends. 

Despite the clear improvements in predictive performance and the computational efficiency of fGFPCA, a few open questions remain. In both the simulation study and data application, the tuning parameter for the fGFPCA approach, bin width, was chosen arbitrarily. In reality, such choices can affect the predictive performance and sensitivity to the choice of bin should be assessed in practice. If the bins are very narrow local GLMMs may be nonidentifiable, resulting in bins with no point estimates. We expect this to happen when bins have a very large proportion of only $0$ or $1$ observed for binary data. Moreover, the current methodology has not been extended to outcomes measured on a sparse or irregular grid, though this is possible in principle. Other methods for GFPCA mentioned in Section \ref{sec:introduction} may be sufficient to deal with these kinds of small, sparse data. One could then use the dynamic prediction framework presented in the current work. 

While fGFPCA showed an advantage in computational efficiency over existing methods based on GLMM, the GLMM framework has the benefit of being able to easily add covariates to the model (e.g., demographic information). An extension of the fast GFPCA methodology used here to covariate-adjusted models is under active development, and we expect our dynamic prediction framework to directly apply once the methodology has been developed for model estimation. 

Despite the limitations of the current work, our proposed method works fast and well for the type of data it was designed for: large, high-density functional data. The current work also opens a number of exciting avenues for future work on different types of functional data. For example, if multiple functions are collected for each individual (e.g., multiple days of minute-by-minute physical activity record, where each day is a single function), then a multilevel GFPCA framework may be more appropriate. We are actively developing a similar prediction framework for multilevel generalized functional data using recent work extending fGFPCA to multilevel data. Another potential future direction is the dynamic prediction of multivariate functional data. Here, different functional outcomes are collected from the same subjects, which are often correlated with each other. For example, activity count and heart rate can be measured by wearable devices at the same time. The future measures of one function can be predicted not only by its own past record but also by the other functions. We leave these exciting extensions as a promising area of future research.

\bibliography{refs}


\newpage

\renewcommand{\thesection}{Supplement \Roman{section}:}
\renewcommand{\thefigure}{S.\arabic{figure}}
\setcounter{section}{0}
\setcounter{figure}{0}
\section{Prediction interval}

The Figure \ref{fig:sim_ci} below presents the coverage rate of the 95\% prediction interval along the unobserved track. The prediction interval of fGFPCA is derived from the pointwise sampling quantiles of the estiamted latent function at each measurement point. On the other hand, the prediction intervals of the other three methods other are Wald intervals calculated by the estimated standard deviation. Coverage is averaged pointwise across all simulated datasets, and the color of curves indicates different length of observed track. Figure \ref{fig:simN500_ci} on the top panel visualizes the output of the first simulation scenario, where training sample size is 500 and the GLMM model is linear. Figure \ref{fig:simN100_ci} on the bottom panel visualizes the output of the second simulation scenario, where training sample size is 100 and the GLMM model incorporates spline basis functions.

Comparatively, fGFPCA achieves the highest coverage rate across the predicted track among the four models, and the coverage rate improves as the observed track extends. However, it did not reach the nominal value of 95\% for a large proportion of the track. The other three methods seem to severly underestimate the uncertainty of prediction, causing the coverage rate to be close to zero. GLMM model has a better coverage when with greater model flexibility, but still not as well as the fGFPCA intervals. 

Improvement of the coverage rate of prediction interval from fGFPCA is an interesting topic for future research. The current interval is a posterior estimate conditional on parameters estimated from fGFPCA algorithm, thus fails to take into account the variability from the population-level estimates from fGFPCA algorithm. This is likely the cause of the coverage rate being lower than nominal level. However, quantiying such variation is a non-trival problem, especially the variation of estimated functional principal components. 

\begin{figure}
    \centering
    \begin{subfigure}{\textwidth}
    \includegraphics[width=\textwidth]{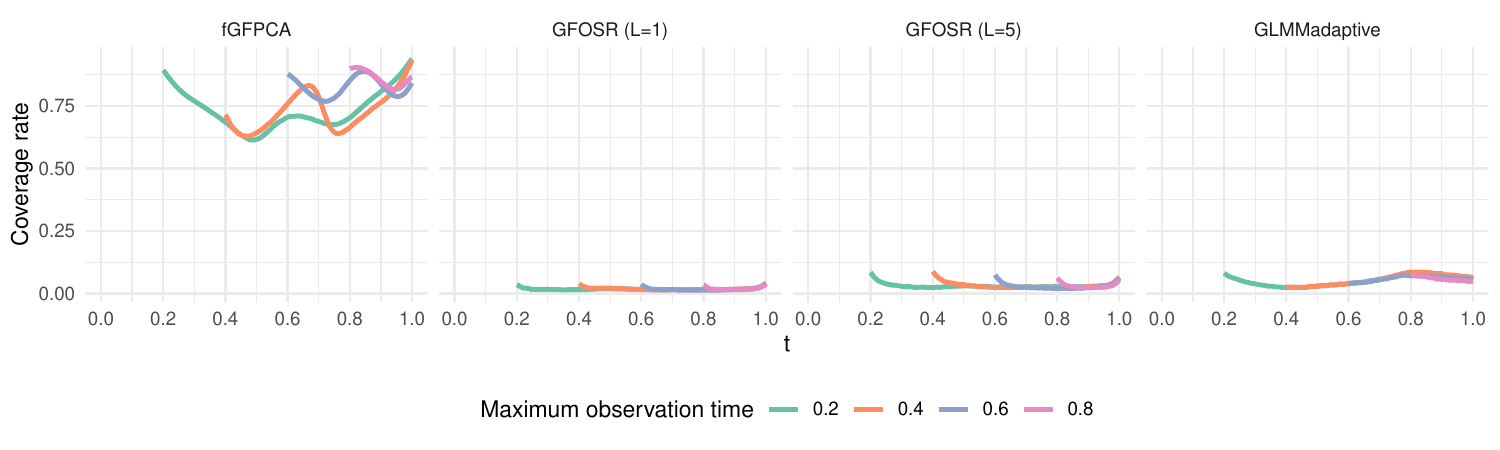}
    \caption{Training sample size = 500}
    \label{fig:simN500_ci}
    \end{subfigure}

    \begin{subfigure}{\textwidth}
    \includegraphics[width=\textwidth]{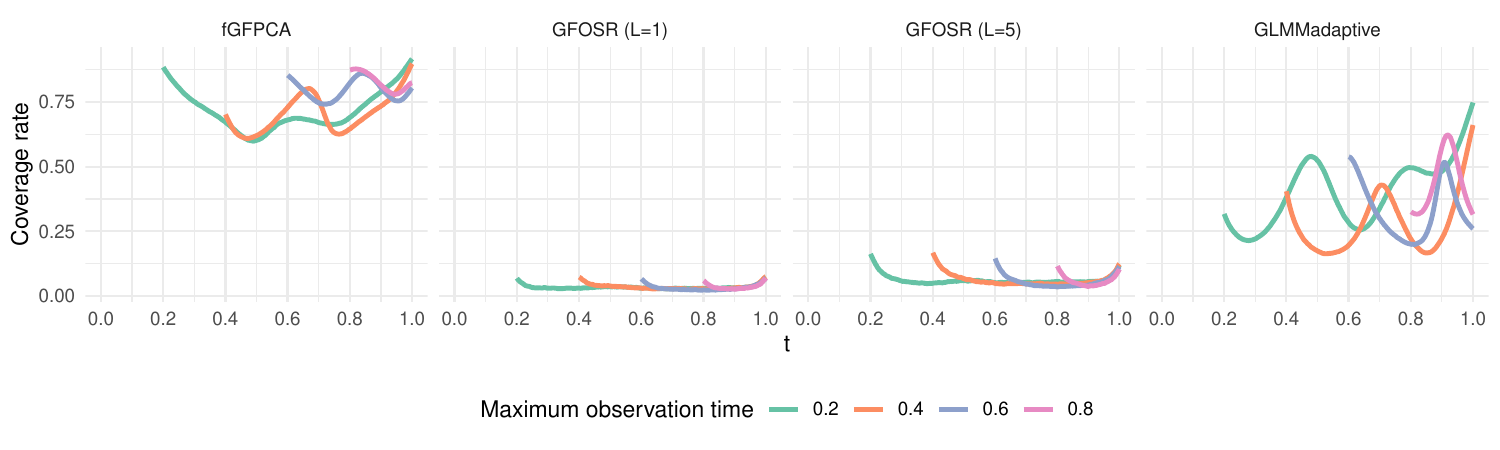}
    \caption{Training sample size = 100}
    \label{fig:simN100_ci}
    \end{subfigure}

    \caption{Converage rate of prediction interval averaged across all simulated datasets without numeric issue.}
    \label{fig:sim_ci}
    
\end{figure}

\end{document}